\documentclass[aps,pre,showpacs,twocolumn,groupedaddress,amssymb,amsmath,10pt,nofootinbib]{revtex4-1}

\usepackage{graphicx}
\usepackage{subfigure}
\usepackage{color}

\newcommand{\beq}{\begin{equation}}
\newcommand{\eeq}{\end{equation}}
\newcommand{\beqn}{\begin{eqnarray}}
\newcommand{\eeqn}{\end{eqnarray}}
\newcommand{\bsm}{\begin{smallmatrix}}
\newcommand{\esm}{\end{smallmatrix}}
\newcommand{\ndelta}{\bar{\delta}}
\newcommand{\tG}{\tilde{\Gamma}}

\begin{document}


\title{Mobility-Dependent Selection of Competing Strategy Associations}


\author{Alexander Dobrinevski$^{1}$}
\author{Mikko Alava$^{2}$}
\author{Tobias Reichenbach$^{3}$}
\author{Erwin Frey$^{4}$}
\affiliation{$^{1}$CNRS-Laboratoire de Physique Th\'eorique de l'Ecole
Normale Sup\'erieure, 24 rue Lhomond, 75005 Paris
Cedex-France\\
$^{2}$Aalto University, School of Science, Department of Applied Physics, PO Box 11100, 00076 Aalto, Finland\\
$^{3}$Department of Bioengineering, Imperial College London,  South Kensington Campus, London, SW7 2AZ \\
$^{4}$Arnold Sommerfeld Center for Theoretical Physics and Center for NanoScience, Department of Physics, Ludwig-Maximilians-Universit\"{a}t M\"{u}nchen, Theresienstrasse 37, D-80333 M\"{u}nchen, Germany
}

\date{\today}

\begin{abstract}
Standard models of population dynamics focus on the the interaction, survival, and extinction of the competing species individually. Real ecological systems, however, are characterized by an abundance of species (or strategies, in the terminology of evolutionary-game theory) that form intricate, complex interaction networks. The description of the ensuing dynamics may be aided by studying associations of certain strategies rather than individual ones. Here we show how such a higher-level description can bear fruitful insight. Motivated from different strains of colicinogenic \emph{Escherichia coli} bacteria, we investigate a four-strategy system which contains a three-strategy cycle and a neutral alliance of two strategies. We find that the stochastic, spatial model exhibits a mobility-dependent selection of either the three-strategy cycle or of the neutral pair. We analyze this intriguing phenomenon numerically and analytically.
\end{abstract}


\maketitle

\section{Introduction\label{sec:Introduction}}

Ecological systems are complex assemblies of many interacting species, subspecies, and subtypes~\cite{Smith,Neal,Wright}. \emph{Escherichia coli} bacteria can, for example, produce certain toxins that kill other \emph{E. coli} strains that are sensitive to these toxins. Bacteria can also be resistant to one or more toxins and hence survive the encounter with the toxin producers. Many different such strains, each of which  produces a certain set of toxins, is resistant to those and others, and sensitive to the remaining ones, coexist in nature.

Recent experimental and theoretical research has focussed on the coexistence of strains that emerge from a single toxin. Bacteria can then be toxin producers (P), sensitive to the toxin (S), or resistant (R). As stated above, the toxin-producing bacteria kill the sensitive ones. Because resistance is mediated by a plasmid that limits nutrient uptake and hence slows reproduction, the sensitive bacteria outgrow the resistant ones. Producing a toxin is an additional metabolic cost, and the bacteria that follow that strategy hence reproduce even slower than the resistant ones. The three strains in question hence exhibit a cyclic competition---reminiscent of the children's game rock, paper, scissors---in which each strain outperforms another but loses against the remaining one. Three such strains can coexist as long as they can spatially organize into different dynamic domains~\cite{durrett-1997-185,durrett-1998-53,czaran-2002-99,Kerr2002,Reichenbach2007,Szabo2007,RulandsZielinskiFrey2013}. When spatial separation is impeded by mixing, two of the three strategies typically extinct rapidly.

The structure and dynamics of real microbial communities, such as biofilms, display a richness of inter-species competitive interactions that goes far beyond the simple rock-paper-scissors paradigm~\cite{Rao2005,Hibbing2009}. As an example, consider two bacterial toxins. Because bacteria can be producers of each toxin as well as resistant or sensitive to each, we obtain nine distinct strains. We denote by RP a strain that is resistant to the first toxin and produces the second; the remaining strains are designated analogously.  The interaction network of these strategies is quite complicated and contains several three-strategy subcycles~\cite{SzaboCzaran2001a}. As another motif, neutral alliances of two strategies in which neither has a competitive advantage over the other appear. As an example, the strategies RP and PR form such an alliance if the rates for killing and growth regarding the two toxins are equal.

\begin{figure}[b]
\subfigure[~Interaction scheme]{
\includegraphics[width=0.4\columnwidth]{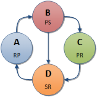}
\label{fig:IntScheme}
}
\subfigure[~Reaction equations]{
\parbox[b][][b]{0.4\columnwidth}{
\beqn
\nonumber A+B  & \stackrel{k_{AB}}{\longrightarrow} A+A , \\
\nonumber B+C & \stackrel{k_{BC}}{\longrightarrow} B+B , \\
\nonumber C+D & \stackrel{k_{CD}}{\longrightarrow} C+C , \\
\nonumber D+A & \stackrel{k_{DA}}{\longrightarrow} D+D , \\
\nonumber B+D &  \stackrel{k_{BD}}{\longrightarrow} B+B .
\eeqn
}
\label{fig:Reactions}
}
\caption{(Color online) The four-strategy system. (a) Four strategies $A$, $B$, $C$, and $D$ compete in a cyclic manner. Because $B$ also dominates $D$, a three-strategy cycle emerges between $A$, $B$, and $D$.  The two strategies $A$ and $C$ form a neutral alliance. Such interactions can arise for colicinogenic bacteria of types R, P, or S (text). (b) Reactions between individuals of the four strategies that produce the required interaction scheme. }
\end{figure}
In the present study, we consider how the simultaneous presence of cyclic dominance and defensive alliances shapes the population dynamics. To this end we focus on four of the nine possible bacterial strategies that emerge for two toxins, namely on RP, PS, PR, and SR. For ease of reference, we abbreviate these strategies as $A$, $B$, $C$, and $D$, respectively (Figure~\ref{fig:IntScheme}). These four strategies then exhibit cyclic dominance (between $A$, $B$, $D$ as well as between $A$, $B$, $C$, $D$) and a neutral alliance (between $A$ and $C$).

Two simple possible steady states are accordingly conceivable. First, the three-strategy cycle $A$, $B$, $D$ can lead to a self-organizing dynamic pattern such as rotating spiral waves~\cite{Kerr2002,Kirkup2004,Reichenbach2007,Reichenbach2008a}.\footnote{{Depending on the diffusion strength, one can also observe systemwide oscillations or convectively unstable spirals \cite{Reichenbach2008,RulandsZielinskiFrey2013}.}} Its stability against intrusion by $C$ remains unclear for strain $C$ can invade strain $D$ but is dominated by $B$.  Second, the two-strategy neutral alliance of strategies $A$ and $C$ yields a static state in which no further dynamics occurs. The formation of such neutral alliances has already been observed and studied~\cite{SzaboCzaran2001,Szabo2004,Szabo2007,Szabo2008,claussen-2008-100,Zia2010,CaseDurneyPleimlingZia2010,he-2010-82,DurneyCasePleimlingZia2011,RomanKonradPleimling2012,dobramsyl-2013-110}. Their robustness against invasion by another strategy association such as a three-strategy cycle, however, remains unclear.

Our study of two competing strategy associations  extends previous discussions of cyclic dominances with four or more strategies \cite{Szabo2004,Zia2010,DurneyCasePleimlingZia2012,RomanKonradPleimling2012,CaseDurneyPleimlingZia2010,DurneyCasePleimlingZia2011,AvelinoEtAl2012,IntoyPleimling2013}.
The $B\to D$ interaction considered here breaks the cyclic symmetry between $A$, $B$, $C$, and $D$, similar to
a recent study of an asymmetric four-strategies interaction network~\cite{LuetzRisauArenzon2012}. In the latter case, varying two of the reaction rates yields  a transition  between a state with all four strategies coexisting and another state with one strategy extinct. Below, we vary the mobility of the individuals which results in two different extinction scenarios.

This article is structured as follows. In the following section \ref{sec:WM}, we consider the model in a well-mixed environment and in the (deterministic) limit of large populations. We show that, except at a critical value of the interaction rates, a certain combination of the interaction strengths determines whether the three-strategy cycle or the neutral alliance survives. We then discuss the importance of stochastic fluctuations at the critical value and for the final extinction process.

In section \ref{sec:Spatial}, we add spatial degrees of freedom. Individuals interact with their nearest neighbors on a two-dimensional lattice on which they are also mobile, leading to local mixing. We identify again the two survival scenarios, namely cyclic dominance  of the three-strategy association $A$, $B$, $D$ as well as the neutral alliance between $A$ and $C$. In contrast to the well-mixed case, each of these strategies is a stable steady state. We find that there is a critical value of the mixing rate, such that for low values of mixing the three-strategy cycle $A$, $B$, $D$  survives, and for high values the neutral alliance $A$, $C$ emerges. We investigate this transition numerically and analytically, using a pair approximation for $2\times 2$ site clusters. Near the transition, we observe behavior similar to a thermodynamical first-order phase transition: the survival probability of the $AC$ neutral alliance changes discontinuously across the transition, and there is no diverging length scale of fluctuations.  We analyze numerically the motion of domain walls between the $A$, $B$, $D$ and the $A$, $C$ domains, as well as the growth of droplets of one strategy association inside the other.

In section \ref{sec:Concl} we summarize our results and discuss their applicability to more general interaction schemes.

\section{The well-mixed environment\label{sec:WM}}

We model the interactions in Figure \ref{fig:IntScheme} through chemical reactions between individuals of the four different strategies (Figure \ref{fig:Reactions}). In a well-mixed environment every individual can potentially interact with every other. A mean-field approach then yields deterministic rate equations for the temporal development of the densities $a$, $b$, $c$, and $d$ of strategies $A$, $B$, $C$, and $D$, respectively~\cite{VanKampen}:
\beq
\partial_t x_i = x_i \sum_{j=1}^4 \left(\Gamma_{ij}-\Gamma_{ji}\right)x_j.
\label{eq:RateEqns}
\eeq
Here we have arranged the strategies' densities in a vector $\vec{x}=(a,b,c,d)$. These densities are the  relative abundances of the strategies $A,B,C$, and $D$, respectively. The matrix $\Gamma$ contains the reaction rates,
\beq
\Gamma=\left( \begin{array}{cccc}
	0 & k_{AB} & 0 & 0 \\
	0 & 0 & k_{BC} & k_{BD} \\
	0 & 0 & 0 & k_{CD} \\
	k_{DA} & 0 & 0 & 0
\end{array} \right).
\label{eq:IntMatrix}
\eeq
Note that the reactions conserve the total number of individuals such that the densities sum to one: $\sum_{i=1}^4x_i=1$. One of the four equations in~(\ref{eq:RateEqns}) is hence redundant, and the phase space is three-dimensional.

The rate equations~(\ref{eq:RateEqns}) have recently been analyzed in detail \cite{KnebelKruegerWeberFrey2013}. The following quantities provide insight into their behavior:
\beqn 
\tau & = & a^{k_{CD}}c^{k_{DA}} ,
\quad
\rho  =  a^{k_{BD}}b^{k_{DA}}d^{k_{AB}} .
\label{eq:RhoTau}
\eeqn
 Recall that $a,b,c,d \in [0;1]$ are the densities or concentrations of the individuals of the different strategies as introduced above. 
These two quantities inform on the presence of the neutral alliance between $A$ and $C$ as well as on that of the three-strategy cycle $A$, $B$, $D$. The first quantity $\tau$ vanishes if and only if the neutral alliance $A$, $C$ disappears. The second quantity, $\rho$, is positive precisely when all three strategies $A$, $B$, and $D$ are found in the population.

A straightforward calculation shows that the equations \eqref{eq:RateEqns} imply the following dynamics for $\tau$ and $\rho$:
\beqn \nonumber
\partial_t \tau & = & -\tau\,b\,(k_{DA}k_{BC}-k_{AB}k_{CD}), \\
\partial_t \rho & = & \rho\,c\,(k_{DA}k_{BC}-k_{AB}k_{CD}).
\label{eq:RhoTauDeriv}
\eeqn
Depending on the sign of the Pfaffian of the interaction matrix $\Gamma$,
\beq
\mathrm{pf}(\Gamma) = k_{AB}k_{CD} - k_{DA}k_{BC},
\eeq
the quantities $\tau$ and $\rho$ hence either grow, remain constant, or decline over time. Indeed, as identified in \cite{CaseDurneyPleimlingZia2010,DurneyCasePleimlingZia2012} and generalized in \cite{KnebelKruegerWeberFrey2013}, the sign of the Pfaffian determines crucial aspects of the dynamics:
\begin{enumerate}
	\item If $\mathrm{pf}(\Gamma)<0$, strategy $C$ dies out rapidly (on a time scale $\propto \log N$), and a neutrally-stable cycle of $A$, $B$, and $D$ remains. Due to stochastic fluctuations, two of the three strategies in this cycle go extinct on a time scale that is proportional to $N$~\cite{Reichenbach2006}. Which species survives is determined by the interaction rates, and only subject to stochasticity in the situation of equal reaction rates~\cite{Berr2009}.
			\item If $\mathrm{pf}(\Gamma)>0$, the strategies $B$ and $D$ die out rapidly (on a time scale $\propto \log N$) and the neutral alliance  of $A$ and $C$ remains. The latter is stable even when stochastic fluctuations are included.
		
		\item If $\mathrm{pf}(\Gamma)=0$, equations \eqref{eq:RateEqns} yields neutrally stable, periodic orbits that are defined by constant values of $\tau$ and $\rho$. Fluctuations due to the stochasticity of the reactions then yield a random walk between these orbits. In other words, the values of $\tau$ and $\rho$ fluctuate. Eventually one of them will vanish, which implies that the system evolves into the neutral alliance $A$, $C$ or the cycle between $A$, $B$, and $D$ on a time-scale proportional to the population size \cite{Reichenbach2006,Dobrinevski2010}. After that, evolution proceeds as in the two cases above. 
		This nontrivial result implies that, for example,  $A$ is never the first strategy to die out and $C$ is never the sole survivor. The splitting probability between the two scenarios is a smooth function of the initial condition.
\end{enumerate}

To summarize, the well-mixed case can be understood though analyzing  the rate equations \eqref{eq:RateEqns}. We find a two-step extinction process. 
First, the system evolves to one of the two states anticipated in the introduction, namely either a neutral alliance of $A$ and $C$ or a three-species cycle $A$, $B$, $D$. While the former is a steady state with perpetual coexistence, the latter leads to extinction of all but one species.  The total extinction process (until only one species, or the non-interacting neutral alliance, is left) takes a time $\propto N$.
As detailed above, the reaction rates in the interaction matrix \eqref{eq:IntMatrix} determine which of these two scenarios occurs.  Their values also inform on which species survive.

\section{Spatially-extended model\label{sec:Spatial}}

A bacterial population typically spreads over an extended spatial region, and  interactions occur only locally. The local range of interactions is, for example, controlled by the mobility of individuals: the more they move around, the larger the area in which they interact within a given time interval.

Spatial segregation of competing strategies can promote biodiversity~\cite{May, czaran-2002-99,hassell-1994-370,Durrett1994,Kerr2002,Reichenbach2007}. Regarding cyclic competition of three bacterial strains, experiments and theoretical models show that spatial segregation stabilizes coexistence of strategies  while in a well-stirred environment, all but one would go extinct~\cite{Kerr2002,Kirkup2004,Durrett1994}. Theoretical studies of such spatially-extended population models have also revealed intriguing phase transitions~\cite{SzaboCzaran2001a,Szabo2004,Szabo2008,LuetzRisauArenzon2012}. In a four-strategies cyclic model, for instance, all four strategies self-organize into spiral-like structures for low mobilities, while large domains with two non-interacting strategies form above a certain critical mobility \cite{Szabo2004}.

We focus on analyzing the interaction scheme \ref{fig:IntScheme} on a two-dimensional lattice. Such an environment is computationally accessible, can be easily visualized, and corresponds well to bacteria growing on two-dimensional surfaces in nature as well as in laboratories (on Petri dishes). We consider a square lattice of  $N=L^2$ sites, each of which is occupied by exactly one individual of the strategies $A$, $B$, $C$, or $D$. An individual can then interact with its four nearest neighbors according to the reactions  detailed in Figure \ref{fig:Reactions}. This means that in a time interval $d t$, an $A$ individual next to a $B$ individual can, for instance, invade the latter with probability $k_{AB}d t$, leaving both sites occupied by $A$ individuals. For simplicity we set all reaction rates to unity: $k_{AB}$, ..., $k_{BD}=1$. We also allow exchange reactions between nearest neighbours at a mixing rate $\epsilon$ that characterizes the mobility of the individuals.  This means that in a time interval $d t$, different individuals occupying neighbouring sites can exchange their places with a probability $\epsilon \,d t$. On a finite lattice, in the limit of continuous time $d t \to 0$, only one of these exchange reactions and the interaction reactions described above may occur at the same time. We thus simulate the stochastic lattice model by randomly choosing the next reaction (among the various interactions and the exchange reaction, taking into account the respective rates) and the nearest-neighbour pair where it occurs (\textit{random sequential updating}).
We use periodic boundary conditions.

\subsection{Numerical results\label{sec:DA}}

\begin{figure}
\flushleft{Low mixing, $\epsilon=0.04$:}\\
\centering
\subfigure[~$t=10$]{
\includegraphics[width=0.14\textwidth]{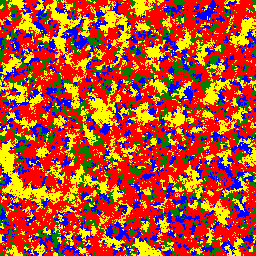}
\label{fig:snapshotLowDiff10}
}
\subfigure[~$t=100$]{
\includegraphics[width=0.14\textwidth]{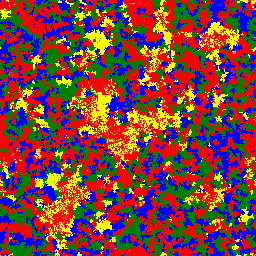}
\label{fig:snapshotLowDiff100}
}
\subfigure[~$t=500$]{
\includegraphics[width=0.14\textwidth]{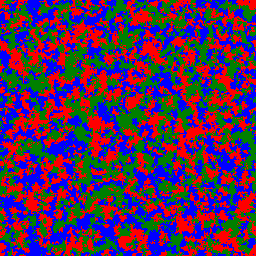}
\label{fig:snapshotLowDiff500}
}
\flushleft{High mixing, $\epsilon=0.08$:}\\
\subfigure[~$t=10$]{
\includegraphics[width=0.14\textwidth]{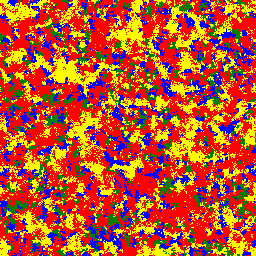}
\label{fig:snapshotHighDiff10}
}
\subfigure[~$t=100$]{
\includegraphics[width=0.14\textwidth]{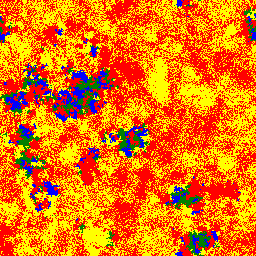}
\label{fig:snapshotHighDiff100}
}
\subfigure[~$t=500$]{
\includegraphics[width=0.14\textwidth]{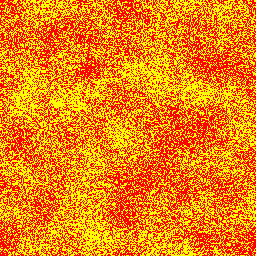}
\label{fig:snapshotHighDiff500}
}
\caption{(Color online) Snapshots of a simulation of the spatially-extended model on a lattice of $256\times 256$ sites at various times $t$. At $t=0$, each site is occupied randomly, with equal probability, with an individual of the four strategies $A$ (red or gray), $B$ (blue or dark gray), $C$ (yellow or light gray), or $D$ (green or medium gray). A low mixing rate then leads to the rapid extinction of strategy $C$ and the prevalence of the dynamic three-strategy  cycle $A$, $B$, $D$. At a higher mixing rate, however, strategies $B$ and $D$ die out, leaving a frozen steady state of the neutral alliance $A$ and $C$. \label{fig:DefAll}}
\end{figure}

Within a well-mixed environment, equality of the reaction rates implies that we are in case 3 [$\textrm{pf}(\Gamma$=0)] of the above analysis (section \ref{sec:WM}). It then takes a long time, proportional to the system size, for one strategy association to go extinct, and which one does is probabilistic.

On a lattice, the extinction process happens much quicker, and the mixing rate determines which strategy association prevails (Figure~\ref{fig:DefAll}).
When the mixing rate is low, strategy $C$ always dies out (in the limit of infinitely large lattices), and the three-strategy cycle $A$, $B$, $D$ remains as a dynamic but stable steady state~\cite{Reichenbach2008,Szabo2004a,RulandsZielinskiFrey2013}. This development resembles scenario 1 of the well-stirred environment.

For large values of $\epsilon$, we observe that strategies $B$ and $D$ always die out, while the non-interacting pair $A$ and $C$ survives. During this process frozen domains of the neutral alliance $A$ and $C$  form and grow. Note that this happens without any explicit attractive interaction between $A$ and $C$. The global dynamics resembles scenario 2 observed in the well-stirred system.

\begin{figure}
\includegraphics[width=0.48\columnwidth]{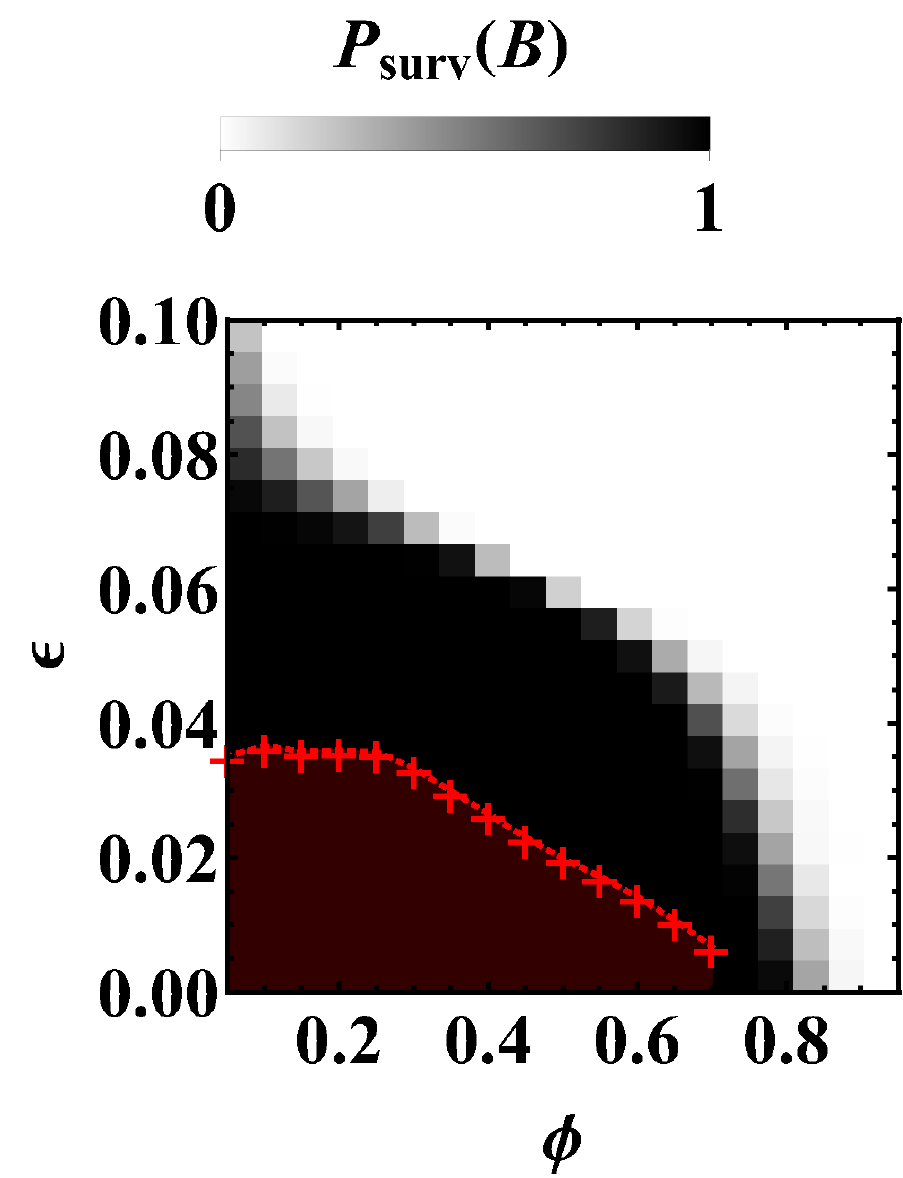}
\includegraphics[width=0.48\columnwidth]{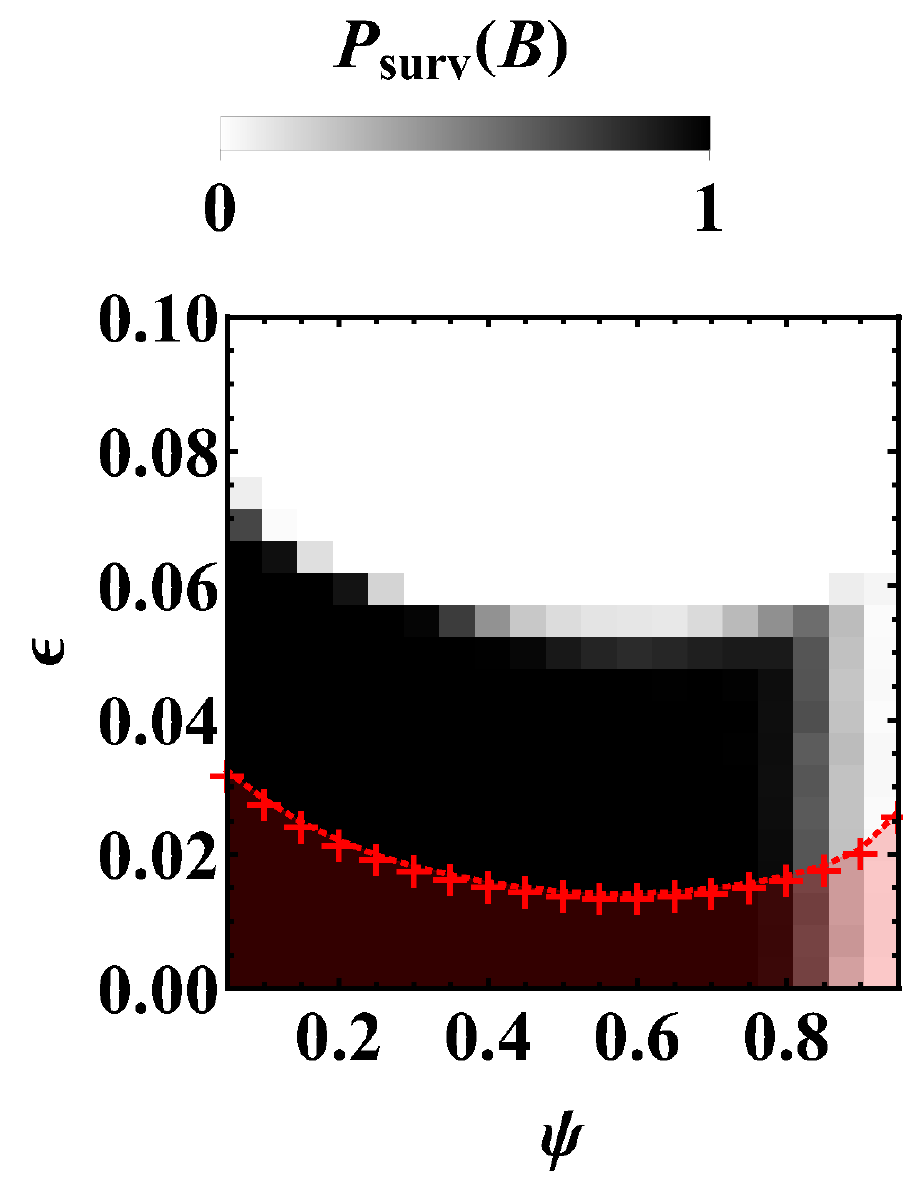}
\caption{(Color online) Phase diagram for the spatially-extended model. We show the survival probability $P_\text{surv}(B)$ of strategy $B$, that informs which strategy association prevails, as a function of the initial densities and the mixing rate. Black hence corresponds to the $ABD$ phase, and white to the $AC$ phase. The initial densities are $\vec{x}=\left(\frac{\phi}{2},\frac{1-\phi}{2},\frac{\phi}{2},\frac{1-\phi}{2}\right)$ (left) and $\vec{x}=\left(\frac{1-\psi}{3},\frac{1-\psi}{3},\psi,\frac{1-\psi}{3}\right)$ (right). Lattice simulations are from a grid of  $N=256\times 256$ sites. Results  for the critical value $\epsilon_c^\text{ana}$ of the mixing rate obtained from a  generalized pair approximation (red crosses)  agree qualitatively, though not quantitatively, with the numerical results. The narrow $AC$ strip for $\psi > 0.8$ in (b) is an artifact of finite system size, and becomes smaller as $N$ increases.}
\label{fig:PhaseTrans}
\end{figure}

The transition between the low- and the high-mobility steady states occurs sharply at a critical mixing rate $\epsilon_c$. Indeed, we can distinguish both steady states through the survival probability of strategy $B$, which vanishes when the steady state contains only the neutral alliance $A$, $C$ but is unity for the three-strategy cycle $A$, $B$, $D$.  Simulation results obtained from averages over many realizations show a sudden transition in $B$'s survival probability (Figure~\ref{fig:PhaseTrans}). For low mixing, the survival probability is unity, indicating that the three-strategy association $A$, $B$, $D$ survives. Large mixing rates lead to an extinction of $B$ and hence the prevalence of the neutral pair $A$ and $C$. The critical value of the mixing rate at which the transition occurs depends on the initial conditions.

The system hence displays a \textit{mobility-dependent selection of strategy associations}. For low mobilities, it is favorable to be part of a rock-paper-scissors-type cycle, whereas for higher mobilities, the neutral alliance takes over.  Coexistence of all four strategies on an extended time scale is only possible at the critical value $\epsilon_c$: extinction is then driven by fluctuations only and requires a time proportional to the system size.

This mobility-dependent selection is highly non-trivial. As an example, once a domain of the neutral pair $A$ and $C$ forms, it can potentially be invaded by strategy $B$ which dominates over $C$. Such a domain, however, can  also defend itself  against a $B$ intruder for strategy $B$  is dominated by  $A$. Which of both scenarios happens depends, as our simulations show, on individual's mobility! Similarly, it may appear intuitive that the critical value of the mixing rate varies with the initial condition for, say, a large initial density of $A$ and $C$ should favor the prevalence of this neutral alliance. The precise form of this dependence, however, is much more difficult to infer intuitively. In the following section we study analytical approximations to provide insight into these issues.

\subsection{Generalized pair approximation\label{sec:GPA}}

In order to gain an analytic understanding of the mobility-dependent phase transition and its dependence on the initial condition, we utilize a generalized mean-field approximation for $2\times 2$ site clusters~\cite{Szabo2004a,Szabo2007}.

Let us summarize the four strategies in a vector $\vec{s}=(A,B,C,D)$, and denote  the probability for a site to be occupied by strategy $s_i$, $i=1,...,4$, as $p^{(1)}(s_i)$. The evolution of these probabilities is given by a master equation that involves the pair probabilities $p^{(2)}(s_i,s_j)$ to have nearest neighbors $s_i$ and $s_j$:
\beq
\partial_t p^{(1)}(s_i) = \sum_{j=1}^4 \left[\tG_{ij}  - \tG_{ji}\right]  p^{(2)}(s_i,s_j).
\label{eq:GPAMasterEqn}
\eeq
The matrix  $\tilde{\Gamma}$ contains the reaction as well as the mixing rates: $\tG_{ij} = \Gamma_{ij} + \epsilon(1-\delta_{ij})$.

In the mean-field approximation one  neglects correlations between a site and its neighbours, and hence assumes that $p_2(s_i,s_j)=p_1(s_i) p_1 (s_j)$ which leads to
\beq
\partial_t p^{(1)}(s_i) = p^{(1)}(s_i)  \sum_{j=1}^4\left[ \tG_{ij}  -  \tG_{ji} \right] p^{(1)}(s_j).
\eeq
Because the probability $p^{(1)}(s_i)$ of strategy $s_i$ is simply that strategy's  density $x_i$, the above equations are equivalent to the rate equations \eqref{eq:RateEqns}.

Describing spatial effects, and hence correlations, requires to go beyond the mean-field approximation. The simplest extension is to consider pair correlations. We have found that those are still not sufficient for an effective description of our system, and hence investigate  $2\times 2$ clusters of neighboring sites whose probability we denote by $p^{(4)} ( \bsm s_i & s_j \\ s_k & s_l \esm )$.
The temporal development of these probabilities follows from a master equation that involves the probabilities $p^{(6,h)}$ and $p^{(6,v)}$ of, respectively,  horizontally or vertically oriented $2\times 3$ clusters.  The resulting equation is lengthy and detailed in Appendix~\ref{app:master_eq}, Equation~(\ref{eq: master_p4}).

In order to obtain a closed system of equations for the $2\times 2$-cluster probabilities, we impose the following closure for the probabilities of the $2\times3$ clusters:
\beqn \nonumber
p^{(6,h)} (\bsm s_i& s_j& s_k \\ s_l & s_m & s_n \esm) &=& \frac{p^{(4)} (\bsm s_i & s_j \\ s_l & s_m \esm)p^{(4)}(\bsm s_j & s_k \\ s_m & s_n \esm)}{p^{(2)} (s_j,s_m)}\\
\label{eq:GPAClosure}
p^{(6,v)} \left(\bsm s_i & s_j\\ s_k& s_l\\ s_m& s_n \esm\right) &=& \frac{p^{(4)} (\bsm s_i & s_j \\ s_k & s_l  \esm)p^{(4)}(\bsm s_k & s_l \\ s_m & s_n \esm)}{p^{(2)} (s_k, s_l)}
\eeqn
We obtain a system of $4^4=256$ coupled ordinary differential equations. Solving this numerically for a fixed initial condition, we can qualiatively reproduce the simulation results.  For a low mixing rate  the density $c$ of strategy $C$ tends to zero with progressing time, and the densities $a$, $b$,  and $d$ oscillate periodically.  At high mixing, in contrast, the densities $b$ and $d$ vanish quickly whereas $a$ and $c$ approach constant values. Both scenarios are separated by a critical value $\epsilon_c^\text{ana}$ for the mixing rate that agrees qualitatively, though not quantitatively, with the critical value $\epsilon_c$ obtained from numerical simulations (Figure~\ref{fig:PhaseTrans}). In particular, it captures the surprising fact that sometimes, increasing the initial density of strategy $C$ favors the $A$, $B$, $D$ cycle (Figure~\ref{fig:PhaseTrans}(b), around $\psi=0.8$). Note that to obtain the analytical threshold $\epsilon_c^{\text{ana}}$, one needs the full time-dependent solution of the ODE system, starting from the chosen initial condition, and for various values of $\epsilon$. It does not seem possible to determine which of the many stationary states is reached without tracking the evolution of the system through the initial transient (appendix \ref{app:master_eq}).

\section{Nucleation and domain growth}

\begin{figure}[t]%
\includegraphics[width=40mm]{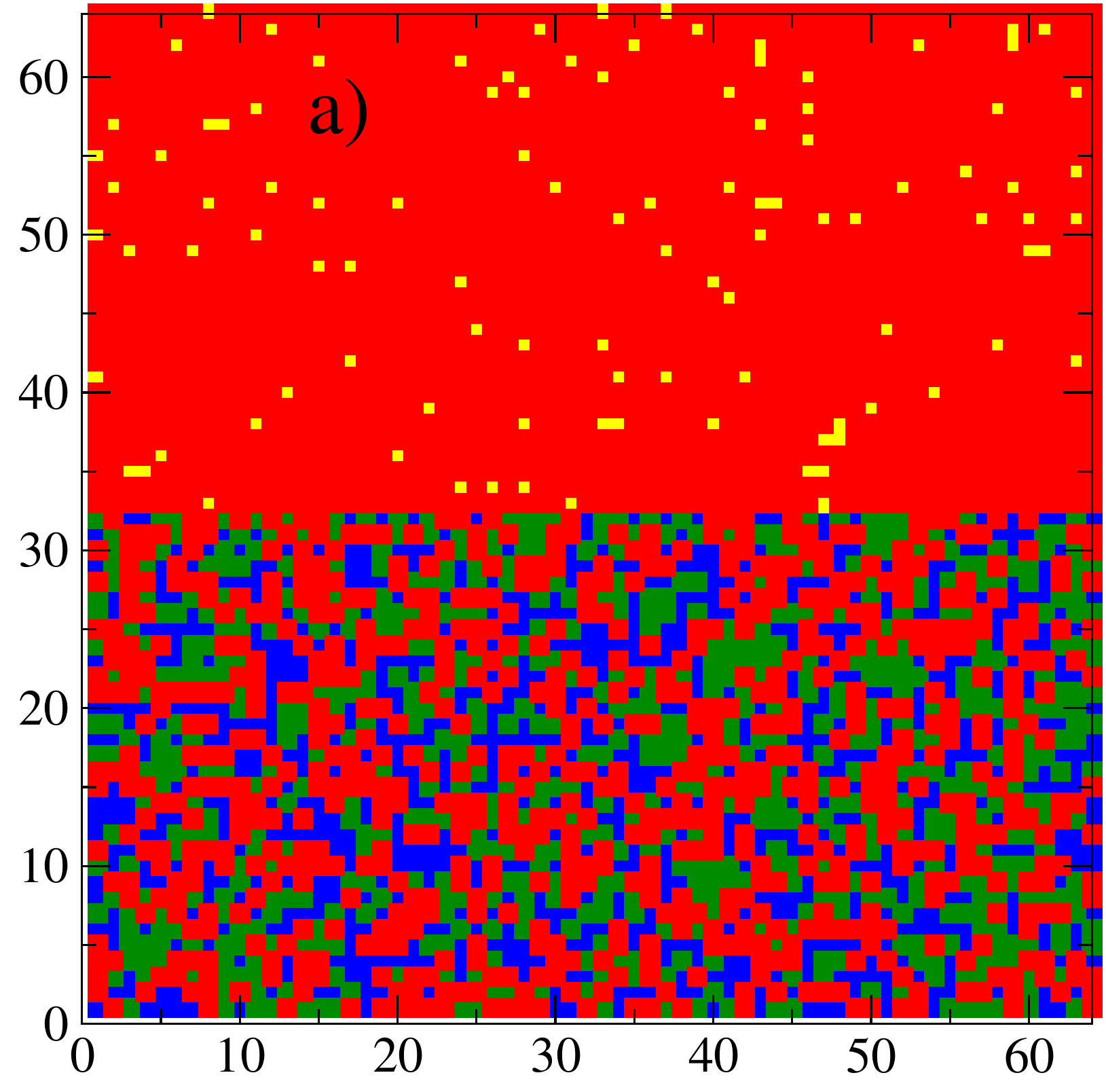}
\includegraphics[width=40mm]{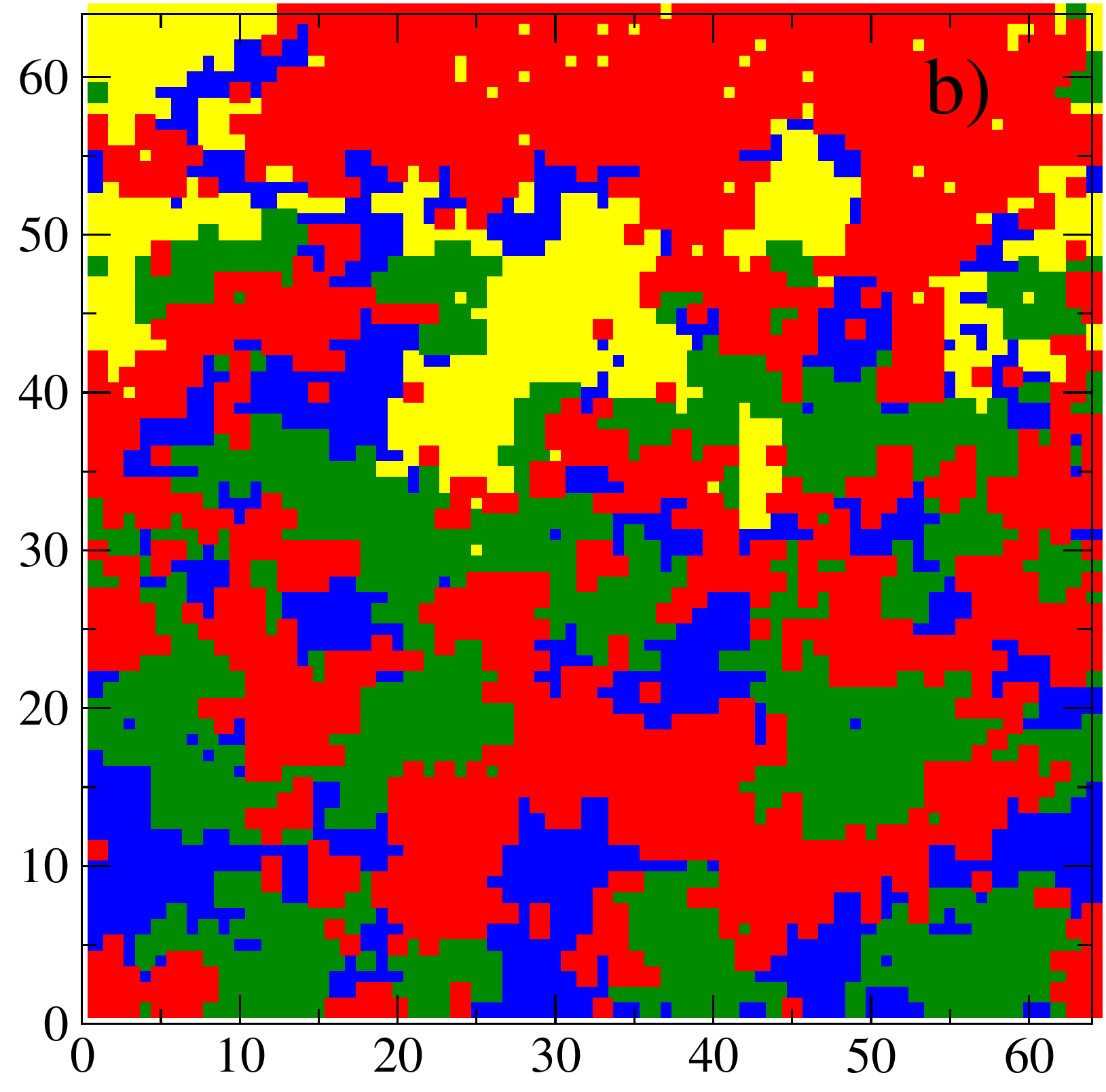}\\
\includegraphics[width=40mm]{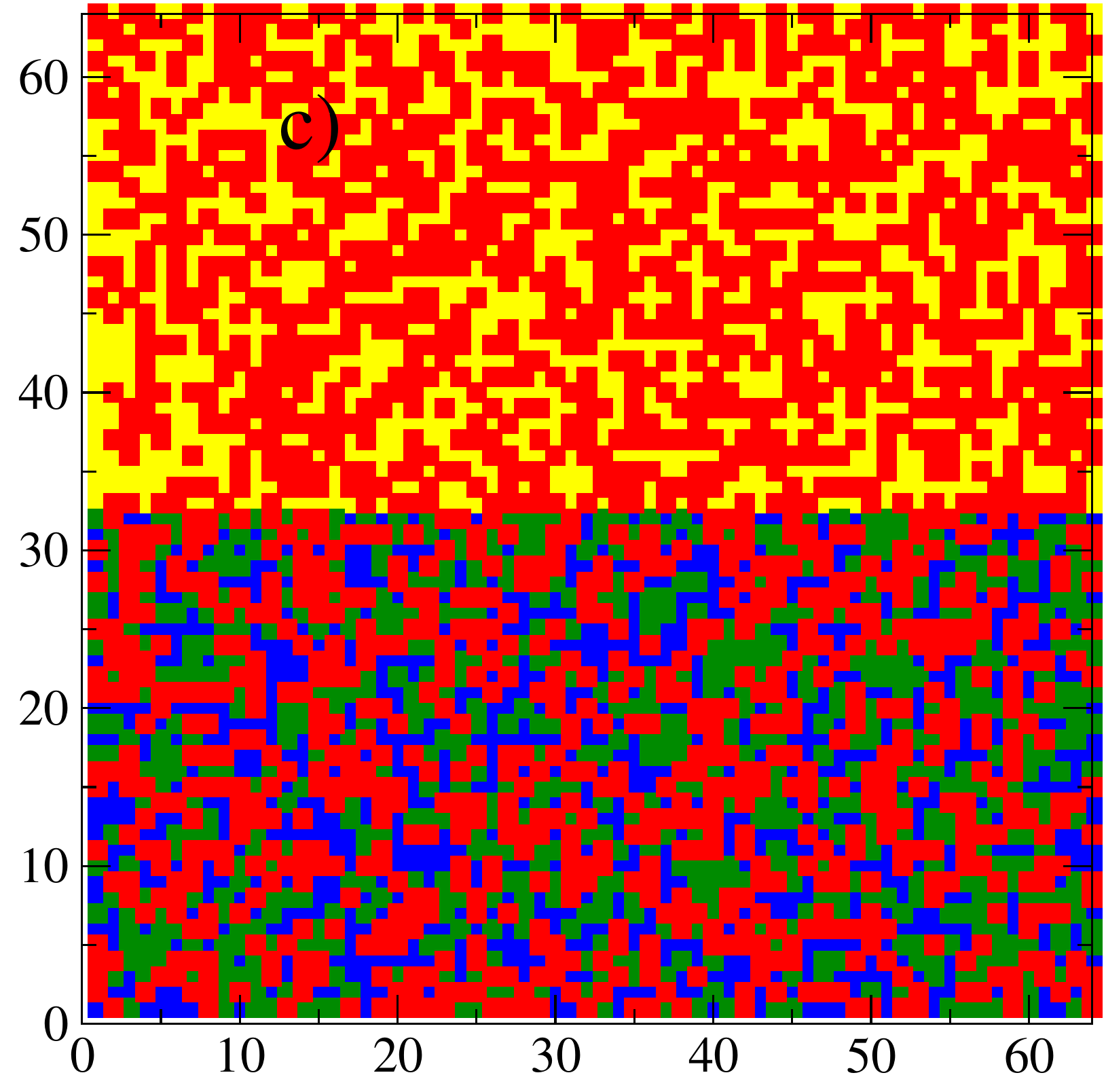}
\includegraphics[width=40mm]{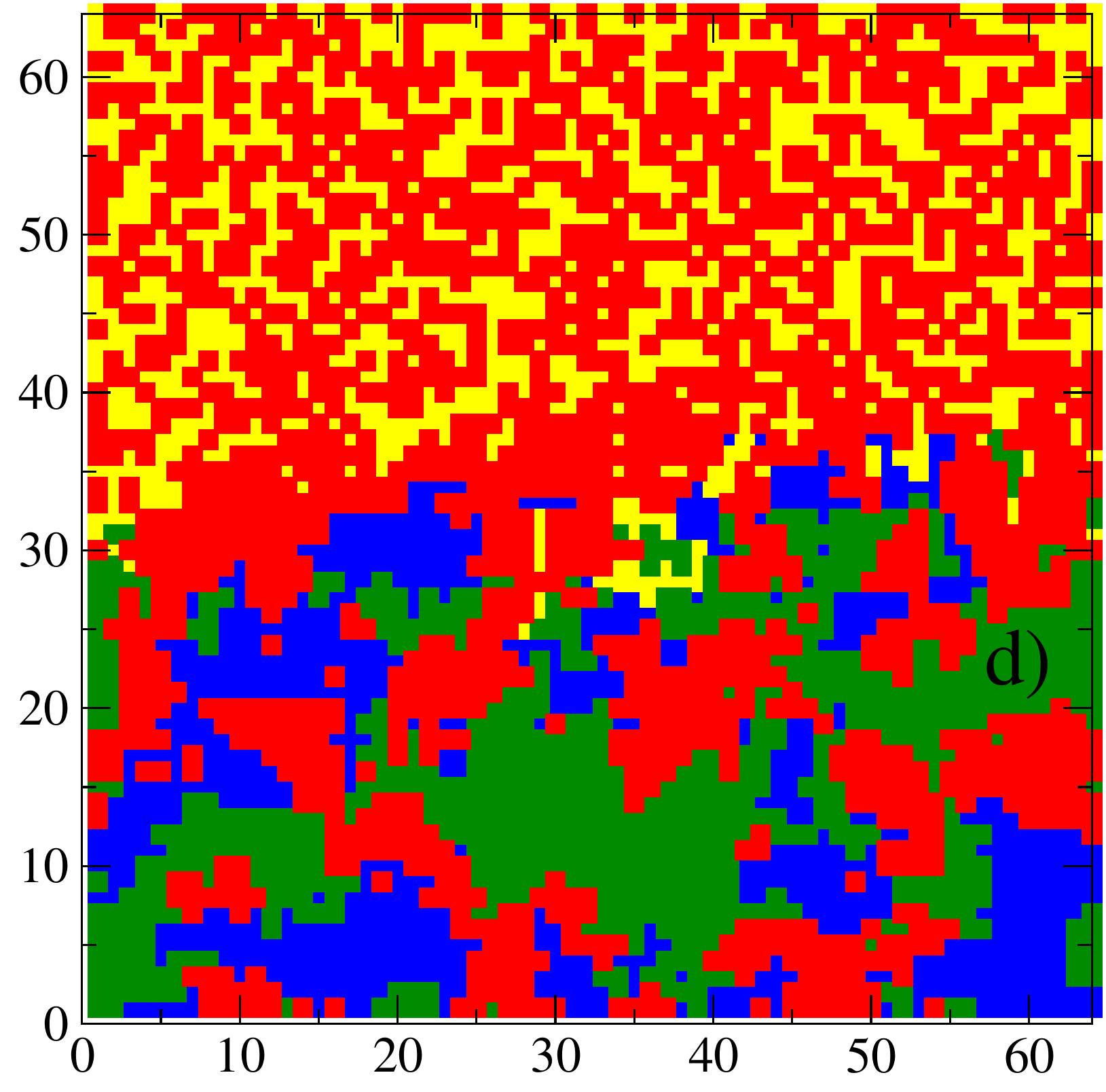}
\caption{(Color online)  Competing domains. (a), (c), Initially, an $A$, $B$, $D$ domain occupies the lower half and an $A$, $C$ domain the upper half. The ratio $a/c$ within the $A$, $C$ domain is $95/5$ in (a) and $1/2$ in (c). (b), (d) The system's state after evolving for $25$ (a) respectively $50$ (b) Monte-Carlo steps. The lattice size is $L=64$, and the mixing rate $\epsilon = 0.03$. }
\label{fig:DWsnap}%
\end{figure}
The analysis of the previous section shows that the non-equilibrium phase transition between the three-strategy cycle and the neutral alliance can be understood on the
basis of short-range correlations (on the order of several lattice sites). Indeed, our simulations show that the system decomposes into growing domains of either the $A$, $C$ neutral alliance or of the  $A$, $B$, $D$ cycle. Some of these regions grow until a single domain occupies the whole system which has then reached its steady state.   Both possible steady states are absorbing for they involve extinction of at least one of the four strategies which cannot reappear.  The phase transition hence
resembles an equilibrium first-order phase transition without a divergent correlation length. A second-order phase transition, in contrast, would involve a steady state in which both domain types coexist, and the average size of one domain type would diverge upon approaching the critical point.

Before a steady state has been reached, domain walls separate the different domains from each other. What is their dynamics, and how does it inform on the system's behavior?

Consider  the motion of a domain wall between a three-strategy domain of  $A$, $B$, and $D$ and an $A$, $C$ cluster.
Let us start from initial conditions such that the lower half of the lattice is filled with a random mixture of $A$, $B$, and $D$ [Figure~\ref{fig:DWsnap} (a), (c)]. A random, frozen mix of $A$ and $C$ individuals occupies the upper half. The $A$, $B$, $D$ domain initially organizes internally, forming patches of the three species. It can then invade the $A$, $C$ domain quickly [Figure~\ref{fig:DWsnap} (b)], slowly  [Figure~\ref{fig:DWsnap} (d)], or be invaded itself (not shown).

\begin{figure}%
\includegraphics[width=\columnwidth]{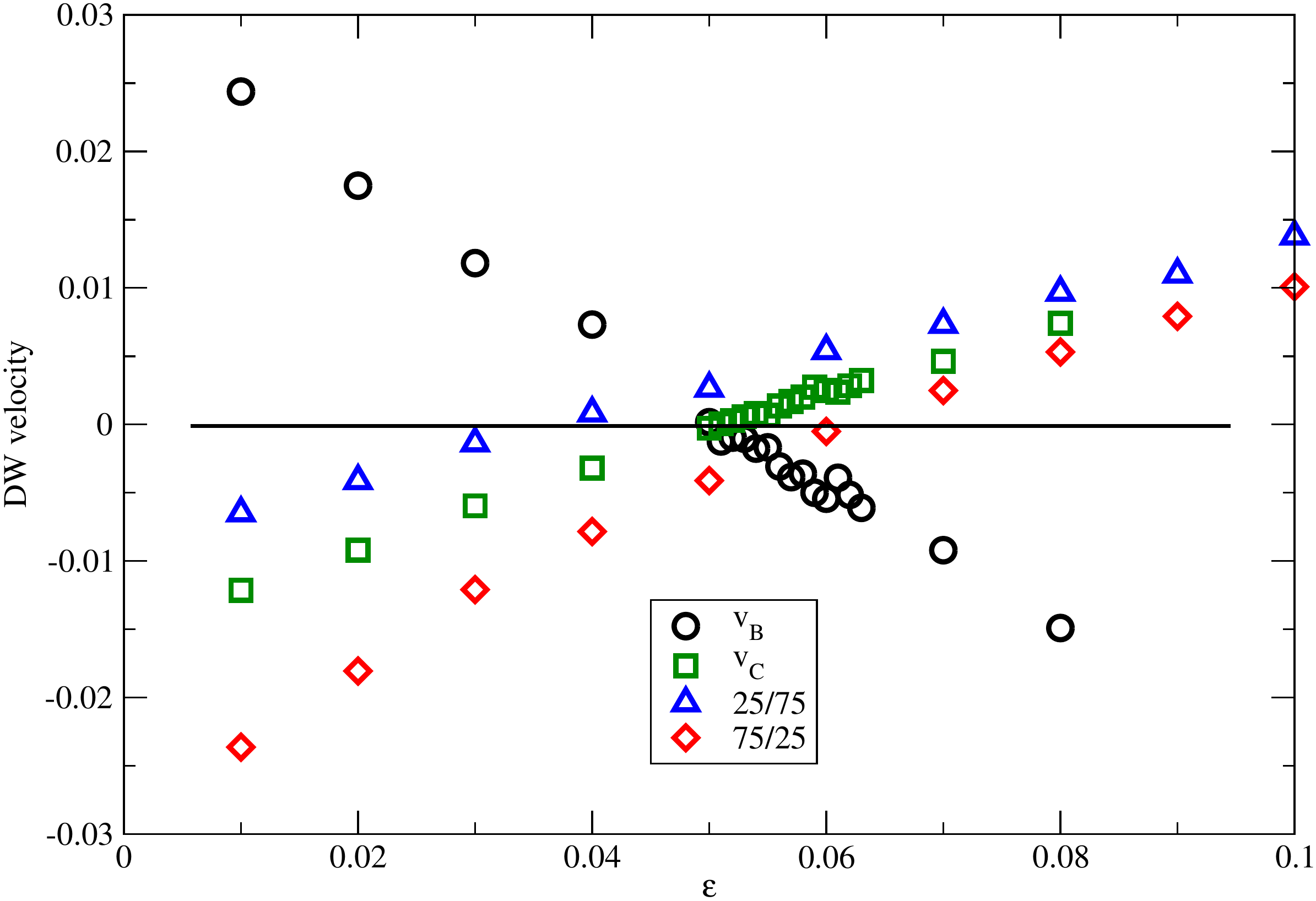}%
\caption{(Color online) Average domain wall velocity.  The numerical results have been obtained from a lattice of linear size $L=128$. The domain wall velocity is positive for low mixing rates, crosses zero at a value $\epsilon_0$, and is negative for large mixing. The value of the zero $\epsilon_0$ depends on the initial conditions. We show results for different ratios of $a/c$ in the $A$, $C$ domain, namely for $a/c=25/75$ (blue triangles), for  $a/c=50/50$ (green squares), and for $a/c=75/25$ (red diamonds). We also include the velocity as seen from the perspective of species $B$: $v_\text{B}=(db/dt)/(2b_0)$ when starting with a $a/c=75/25$ (black circles). We observe that it is the inverse of the velocity $v$ as seen from species $C$.}
\label{fig:DWvelo}%
\end{figure}
Because species $C$ inhabitates only one of the two domains, namely the frozen one, the domain wall position is proportional to the density $c$. If space is measured in units of the lattice size, the domain-wall velocity $v$ reads $v=(dc/dt)/(2c_0)$ in which $c_0$ is the initial density of species $C$. 

The domain wall velocity was analyzed in a time window of up to several hundred Monte-Carlo steps  to ensure that a stable velocity were obtained. Averages are typically from more than 1000 runs. Similar approaches have been used recently to study the phase diagram of multi-species models, with interesting finding regarding the species' densities at the interface~\cite{Szabo2004,Vukov2013} and the roughening kinetics of the front~\cite{Roman2013,Durney2012}. Interesting for future studies would be to investigate whether the roughening of the domain wall observed here belong to the Edwards-Wilkinson or the Kardar-Parisi-Zhang universality class.

Our simulations show that the so-quantified domain-wall velocity is positive for small mixing rates $\epsilon<\epsilon_0$, crosses zero at a value $\epsilon_0$,  and is negative for mixing rates above $\epsilon_0$ (Figure~\ref{fig:DWvelo}). 
For low mixing rates, the three-strategy cycle accordingly invades the neutral alliance, whereas it is invaded itself at large mixing rates.  For random initial occupancies in each domain, the value  $\epsilon_0$ found here is 0.0535,  which is in line with the critical value $\epsilon_c$ in the phase diagram of Figure~\ref{fig:PhaseTrans} for occupation densities of $1/4$ for all strategies.

In contrast to an $A$, $B$, $D$ domain, where the density of each strategy always tends to $1/3$ for long times, the ratio  $a/c$  in the bulk of an $A$, $C$ domain remains constant. There exist accordingly  infinitely many types of stable $A$, $C$ domains, with different ratio of the densities. This ratio influences the domain's survival when competing against the $A$, $B$, $D$ cycle.

We exemplify this effect through measuring the domain-wall velocity  for a varying ratio $a/c$ in the initial $A$, $C$ domain. If strategy $A$ is less frequent than $C$, the domain wall velocity decreases, and the zero $\epsilon_0$ for the mixing rate increases (Figure~\ref{fig:DWvelo}). On a qualitative level this agrees with what we have found before for the critical mixing rate $\epsilon_c$  (Figure~\ref{fig:PhaseTrans}) and explains the unexpected dependence of the critical value on the initial condition.

\subsection{Droplet survival}

Finally, we analyze the dynamics of an $A$, $B$, $D$ droplet  embedded in a large $A$, $C$ domain, and vice versa, for various values of $\epsilon$. The system is prepared by inserting a rectangular domain of linear size $R$ into a background of the other domain  (circle-shaped droplets produce similar results).

\begin{figure}%
\includegraphics[width=\columnwidth]{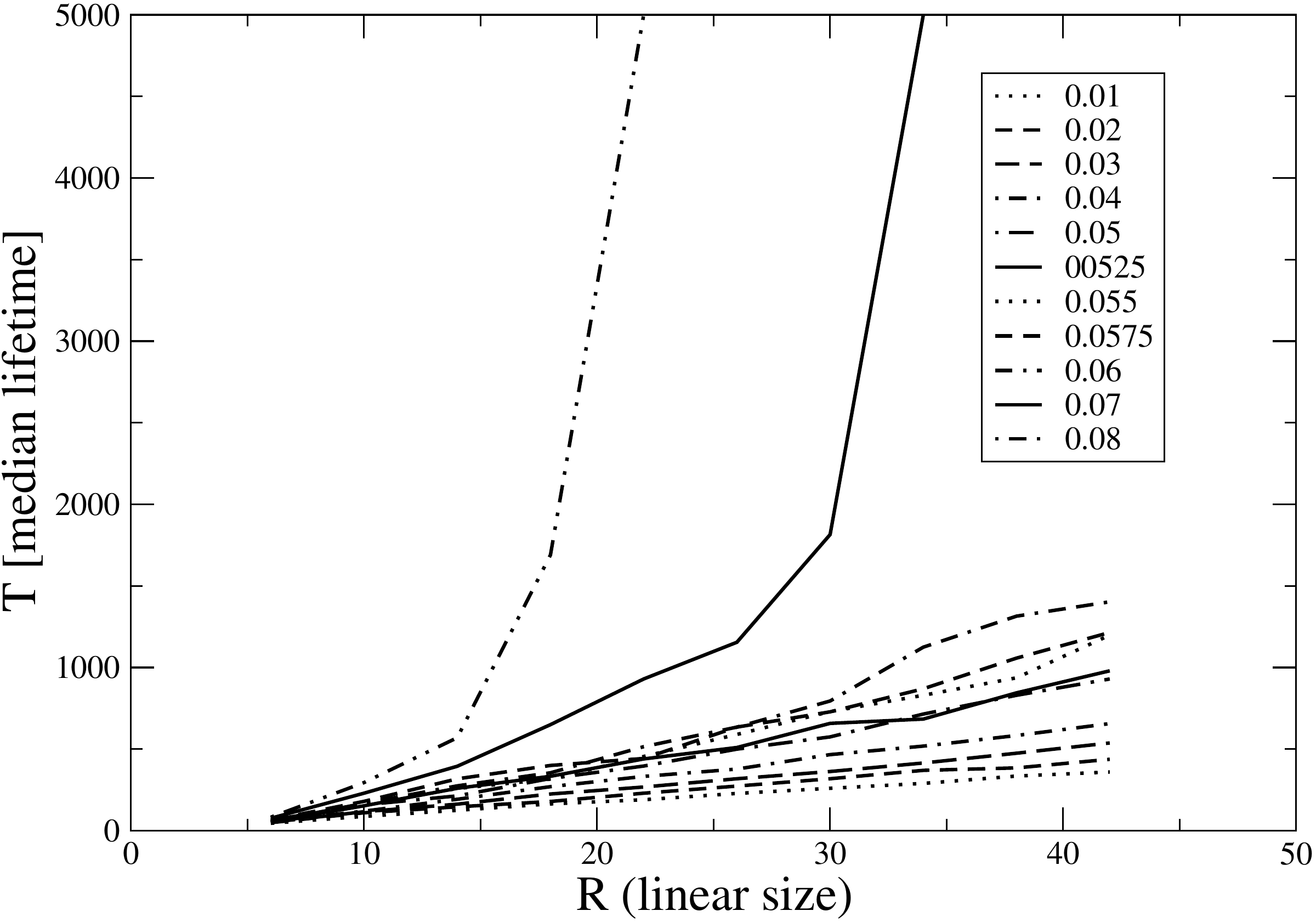}\\
\includegraphics[width=\columnwidth]{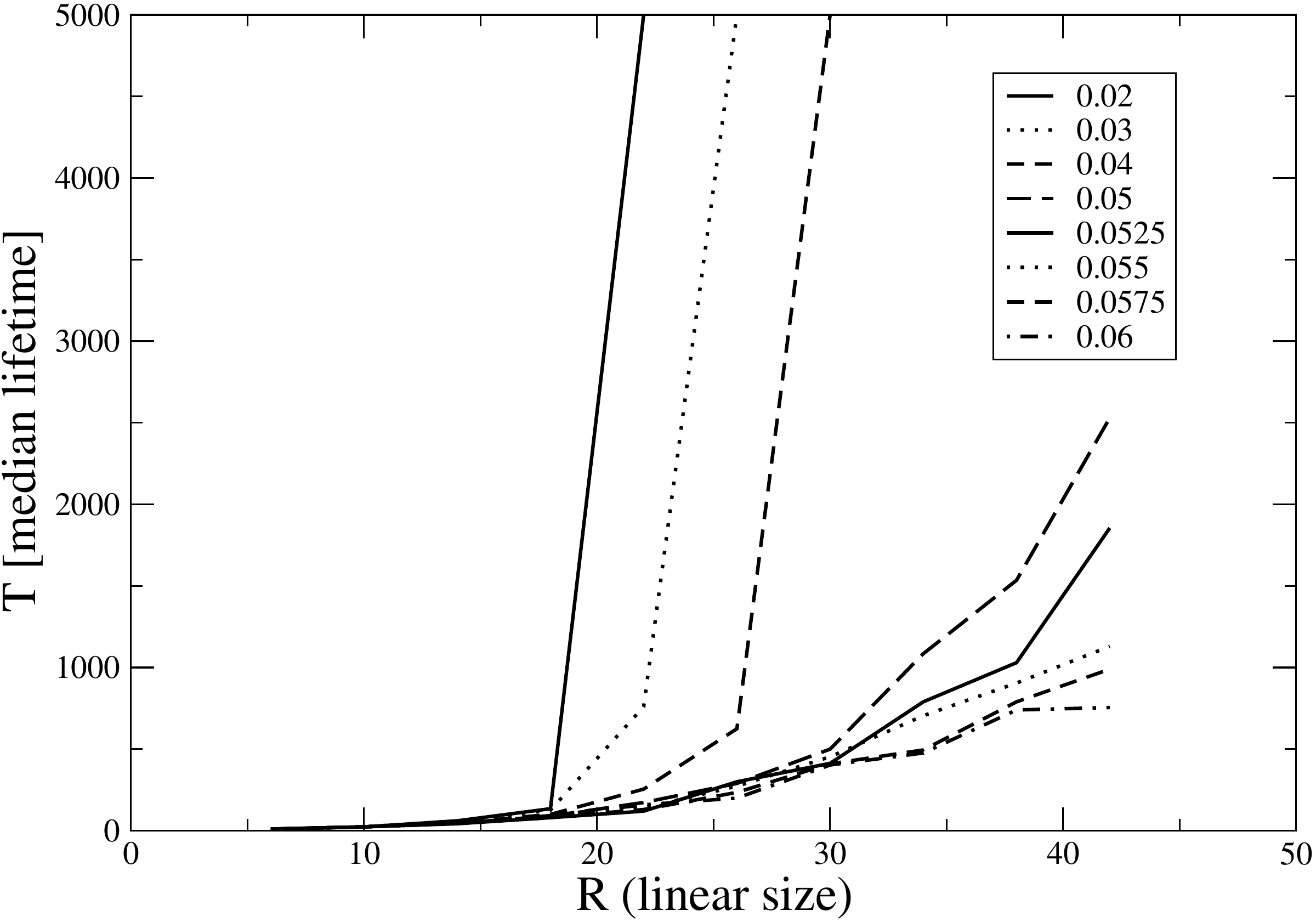}%
\caption{The mean life time $T$  of an $A$, $C$ droplet  (a) and an $A$, $B$, $D$ droplet (b). Results have been obtained from a lattice of linear size $L=128$ and for different values of the mixing rate $\epsilon$ (legends).}
\label{fig:Droplet}%
\end{figure}

We can then monitor the development by following how long, in the situation of an $A$, $B$, $D$ droplet, species $B$ survives.  For small values of the mixing rate, the $A$, $B$, $D$ droplet shrinks, and species $B$ quickly goes extinct. The droplet's mean life time $T$ is then proportional to the initial droplet size, as we expect from a constant domain-wall velocity [Figure~\ref{fig:Droplet}(a)]. For mixing rates above the critical value, however, the droplets expand on average, and extinction takes a very long time.

In the situation of an $A$, $C$ droplet in an  $A$, $B$, $D$ domain, we measure the mean life time for species $C$ [Figure~\ref{fig:Droplet}(b)]. Large values of the mixing rate yield a decay of the droplet size and quick extinction of $C$. Unexpectedly, however, the dependence of the mean extinction time on the droplet size is not linear as in the case of the  $A$, $B$, $D$ droplet. Small mixing rates yield, on average, growing droplets and thus very long extinction times.

\section{Conclusion and Outlook\label{sec:Concl}}

Three-strategy cyclic dominance is a widely used paradigm for explaining biodiversity. It has been shown that this ``rock-paper-scissors'' association is able to sustain coexistence of strategies in a spatially-extended system. In a well-mixed system, two of the three strategies typically go extinct.

Competition does not only occur on the level of individual strategies, however. Although it starts there, it can lead to an effective competition between different strategy associations.
 In this letter, we have discussed such a scenario which emerges in a four-strategy population model. A three-strategy cycle then competes with a two-strategy neutral alliance.   Each of the strategy associations has means of invading and defending against the other.

The main result of our work is that in a spatially-extended population with local interactions, the mobility of the individuals determines which of these two strategy associations wins. For small mobilities, the rock-paper-scissors dynamics is dominant, whereas for large mobilities, the neutral alliance takes over. A state transition that resembles a first-order phase transition occurs at a critical value of the mobility. Near the transition, domains of the three-strategy cycle and the two-strategy neutral alliance evolve and compete.  The average domain-wall velocity changes sign at the critical value of the mobility.

Our numerical results are corroborated  by a generalized pair approximation, which takes into account short-range correlations between neighboring sites. It correctly predicts the dependence of the critical mobility on the initial condition.

To make further progress in the understanding of biodiversity and competition it will be necessary to extend our results to more general interaction schemes~\cite{Zia2010}. In particular, both neutral alliances and dynamic associations, such as the rock-paper-scissors cycle,  are ubiquitous motifs of more complex population dynamics. Their studies will further reveal the importance of studying competition not only on the level of individual strategies but also their associations.

From a statistical physics point of view, it will be highly interesting to further understand the domain formation and growth process (Figure \ref{fig:DefAll}). Its similarities and differences to analogous processes in equilibrium first-order phase transitions merit further studies as well.

\begin{acknowledgments}

We thank Johannes Knebel for useful discussions. Financial support of the German Research Foundation via grant number FR 850/9-1 "Ecology of Bacterial Communities: Interactions and Pattern Formation" is gratefully acknowledged. T.~R. was supported by a Career Award at the
Scientific Interface from the Burroughs Wellcome Fund. M.~A. would like to thank Arnold Sommerfeld Center for Theoretical Physics, Department of Physics, Ludwig-Maximilians-Universit\"{a}t for hospitality, and the Center of Excellence program, Academy of Finland for support. A.D. acknowledges a doctoral fellowship by the CNRS.

\end{acknowledgments}

\appendix

\section{Pair approximation\label{app:master_eq}}

In this appendix we provide more details on the generalized pair approximation, which extends the standard mean-field approach as discussed in section \ref{sec:GPA}. The idea, following Szab\'{o} \cite{Szabo2004,Szabo2007}, is to account for short-range correlations by considering all $4^4=256$ possible $2\times 2$-clusters of neighbouring sites as the underlying states. We denote their probabilities by $p^{(4)} ( \bsm s_i & s_j \\ s_k & s_l \esm )$.
The occupation probabilities $p^{(2)}$ for nearest-neighbour site pairs ($2\times 1$ or $1\times 2$ clusters) and $p^{(1)}$ for single sites ($1\times 1$ clusters) can then be obtained as 
\beqn 
\label{eq:AppP2}
p^{(2)}(s_i,s_j) &=& \sum_{s_k, s_l} p^{(4)}(\bsm s_i & s_j \\ s_k & s_l \esm), \\
\label{eq:AppP1}
p^{(1)}(s_i) &=& \sum_{s_j, s_k, s_l} p^{(4)}(\bsm s_i & s_j \\ s_k & s_l \esm).
\eeqn

The temporal development of the probabilities $p^{(4)}$ of $2\times 2$ clusters follows from a master equation. As usual, it is obtained by enumerating all possibilities of reactions that produce the cluster $( \bsm s_i & s_j \\ s_k & s_l \esm )$ ({\it in} terms), and all possibilities of reactions that alter this cluster ({\it out} terms). 
Since all reactions occur between nearest neighbours, these terms can be written in terms of the probabilities $p^{(6,\text{h})}$ (resp. $p^{(6,\text{v})}$) of horizontal (resp. vertical) $2\times 3 = 6$ clusters:
\widetext{
\beqn \nonumber
\lefteqn{\partial_t p^{(4)} \left(\bsm s_i & s_j \\ s_k & s_l \esm \right)=}&&\\
\nonumber &\sum\limits_{x=1}^4 & \left[ p^{(4)} \left(\bsm s_i & s_x \\ s_k & s_l \esm \right) \Gamma_{ix}\delta_{ij} + p^{(4)} \left(\bsm s_i & s_j \\ s_x & s_l \esm \right) \Gamma_{ix}\delta_{ik}
 + p^{(4)} \left(\bsm s_x & s_j \\ s_k & s_l \esm \right) \Gamma_{jx}\delta_{ji} + p^{(4)} \left(\bsm s_i & s_j \\ s_k & s_x \esm \right) \Gamma_{jx}\delta_{jl}  \right.\\
\nonumber &&\left. + p^{(4)} \left(\bsm s_x & s_j \\ s_k & s_l \esm \right) \Gamma_{kx}\delta_{ki} + p^{(4)} \left(\bsm s_i & s_j \\ s_k & s_x \esm \right) \Gamma_{kx}\delta_{kl}
 + p^{(4)} \left(\bsm s_i & s_x \\ s_k & s_l \esm \right) \Gamma_{lx}\delta_{lj} + p^{(4)} \left(\bsm s_i & s_j \\ s_x & s_l \esm \right) \Gamma_{lx}\delta_{lk} \right] \\
\nonumber + & \sum\limits_{m,x=1}^4 & \left[p^{(6,\text{h})} \left(\bsm s_i & s_x & s_j \\ s_m & s_k & s_l \esm \right)\tG_{ix}
+p^{(6,\text{h})} \left(\bsm s_m& s_i& s_j \\ s_k& s_x& s_l\esm \right)\tG_{kx}
+p^{(6,\text{h})} \left(\bsm s_i& s_x& s_j \\ s_k& s_l& s_m\esm\right)\tG_{jx}
+p^{(6,\text{h})} \left(\bsm s_i& s_j& s_m \\ s_k& s_x& s_l\esm \right)\tG_{lx} \right. \\
\nonumber && \left. +p^{(6,\text{v})} \left(\bsm s_i& s_m \\ s_x& s_j \\ s_k& s_l\esm\right)\tG_{ix}
+p^{(6,\text{v})} \left(\bsm s_m& s_j \\ s_i& s_x \\ s_k& s_l\esm \right)\tG_{jx}
 +p^{(6,\text{v})} \left(\bsm s_i& s_j \\ s_x& s_l \\ s_k& s_m\esm \right)\tG_{kx}
+p^{(6,\text{v})} \left(\bsm s_i& s_j \\ s_k& s_x \\ s_m& s_l\esm \right)\tG_{lx} \right] \\
\nonumber + & \epsilon & \left[\ndelta_{ij} p^{(4)} \left(\bsm s_j & s_i \\ s_k & s_l \esm \right) + \ndelta_{ik} p^{(4)} \left(\bsm s_k & s_j \\ s_i & s_l \esm \right) + \ndelta_{jl} p^{(4)} \left(\bsm s_i & s_l \\ s_k & s_j \esm \right) + \ndelta_{kl} p^{(4)} \left(\bsm s_i & s_j \\ s_l & s_k \esm \right) \right] \\
\nonumber - &\sum\limits_{m,n=1}^4 &\left[ p^{(6,\text{h})} (\bsm s_i& s_j& s_m \\ s_k& s_l& s_n\esm)\left(\tG_{mj}+\tG_{nl}\right)
 +p^{(6,\text{h})} (\bsm s_m& s_i& s_j \\ s_n& s_k& s_l\esm)\left(\tG_{mi}+\tG_{nk}\right)  \right.\\
\nonumber && \left. +p^{(6,\text{v})} \left(\bsm s_i& s_j \\ s_k& s_l \\ s_m& s_n\esm \right)\left(\tG_{mk}+\tG_{nl}\right)
 +p^{(6,\text{v})} \left(\bsm s_m& s_n \\ s_i& s_j \\ s_k& s_l\esm \right)\left(\tG_{mi}+\tG_{nj}\right) \right] \\
\nonumber - &p^{(4)} (\bsm s_i & s_j \\ s_k & s_l \esm) &\left[ \tG_{ij}+\Gamma_{ji}+\tG_{ik}+\Gamma_{ki}
+ \tG_{jl}+\Gamma_{lj} + \tG_{kl}+\Gamma_{lk} \right] .
\label{eq: master_p4}
\eeqn
Here we utilized the shorthand notation $\ndelta_{xy}:=1-\delta_{x,y}$. Lines 1 and 2 contain \textit{in} terms due to interactions inside the cluster. Lines 3 and 4 contain \textit{in} terms that arise from interactions or exchange reactions of the cluster with its neighbours. Line 5 contains \textit{in} terms due to exchange reactions inside the cluster. The remaining lines contain the corresponding \textit{out} terms.

In order to obtain a closed system of equations, we now use the generalized pair approximation \eqref{eq:GPAClosure} to express the $2\times 3$ and $3\times 2$-cluster probabilities $p^{(6,\text{h})}, p^{(6,\text{v})}$ in terms of the $2\times 2$ cluster probability $p^{(4)}$ and the pair probability $p^{(2)}$. The latter can also be obtained from $p^{(4)}$ using \eqref{eq:AppP2} (the choice of sites over which the summation in \eqref{eq:AppP2} is performed is arbitrary and does not influence the result).
Then, the master equation \eqref{eq: master_p4} becomes a system of $4^4=256$ coupled ordinary differential equations for the functions $p^{(4)}(\bsm s_i & s_j \\ s_k & s_l \esm)$. The preparation of the system, where each site is filled independently with species $s=1...4$ with probability $x_s$, yields the initial condition 
\beq
\nonumber
p^{(4)}_0(\bsm s_i & s_j \\ s_k & s_l \esm) = x_{s_i}x_{s_j}x_{s_k}x_{s_l}.
\eeq
We solved the resulting ODE system numerically using Mathematica, and computed the global density of each species $p^{(1)}(s)$ using \eqref{eq:AppP1}.
For long times, one finds the behaviour discussed in section \ref{sec:GPA}. For $\epsilon$ below a critical value $\epsilon_c$, $p^{(1)} (c)$ tends to zero and $p^{(1)} (a)$, $p^{(1)}(b)$, $p^{(1)}(d)$ oscillate periodically, while above a critical value $\epsilon_c$, $p^{(1)}(b)$ and $p^{(1)}(d)$ tend to 0 and $p^{(1)}(a)$ and $p^{(1)}(c)$ approach constant values.  As in the case of the standard RPS game, ref.~\cite{Szabo2004a}, the 2x2 cluster approximation predicts stationary oscillations of the global species densities $a, b,c, d$ in the ABD phase. Similar oscillations are observed when simulating the \textit{finite} lattice model. However, numerical simulations show that these global density oscillations diminish as the lattice size increases; they  are hence a finite-size effect. Regarding our description through coupled ODEs, the oscillations presumably result from the performed approximations, and their amplitude would decrease if one considered larger clusters. 
To  obtain the red curve (GPA prediction) in the phase diagram in figure \ref{fig:PhaseTrans}, we impose a cutoff at $p^{(1)}(a)p^{(1)}(c)=0.001$ and at $p^{(1)}(a)p^{(1)}(b)p^{(1)}(d)=0.001$, respectively, in order to determine when the AC neutral pair or the ABD cycle goes extinct. 
The precise value of this cutoff has a (small) effect on the observed ``extinction time'' at which the cutoff is reached. However, it does not qualitatively affect the outcome, i.e. the type of surviving association, as long as it is smaller than the minimal value reached during the stationary oscillations in the ABD phase or during the initial transient. We checked that this is the case for the points sampled in figure \ref{fig:PhaseTrans}. For determining the location of the transition with very high precision, the cutoff may need to be decreased as one approaches the transition point, since then the transient period is longer and oscillations are more extreme.
We observe that, qualitatively, the phase diagram obtained using the generalized pair approximation agrees well with the one obtained by numerical simulations of the lattice model.

}

 \bibliographystyle{apsrev4-1}


%

\end{document}